\documentclass[letterpaper, 10 pt, conference]{ieeeconf}

\IEEEoverridecommandlockouts
\usepackage{cite}

\usepackage{amsthm,amsmath,amssymb,amsfonts}
\usepackage{algorithmic}
\usepackage[ruled, noline, linesnumbered]{algorithm2e}
\usepackage{graphicx}
\usepackage{xcolor}

\usepackage{array}
\usepackage{multirow}
\usepackage{diagbox}

\makeatletter
\let\MYcaption\@makecaption
\makeatother

\usepackage[font=footnotesize]{subcaption}

\makeatletter
\let\@makecaption\MYcaption
\makeatother

\usepackage{textcomp}
\def\BibTeX{{\rm B\kern-.05em{\sc i\kern-.025em b}\kern-.08em
T\kern-.1667em\lower.7ex\hbox{E}\kern-.125emX}}

\usepackage{lipsum}


\theoremstyle{plain}

\theoremstyle{definition}

\def\({\left(}
\def\){\right)}
\def\[{\left[}
\def\]{\right]}

\def\Gbf{{\bf G}}  

\def\Kbf{{\bf K}}  
\def\Pbf{{\bf P}}  
\def\ubf{{\bf u}}
\def\Wbf{{\bf W}}  \def\wbf{{\bf w}}
\def\Xbf{{\bf X}}  \def\xbf{{\bf x}}
\def\Ybf{{\bf Y}}  \def\ybf{{\bf y}}
\def\Zbf{{\bf Z}}  \def\zbf{{\bf z}}
\def\Phibf{{\bf \Phi}}
\def\Hcal{\mathcal{H}}  
\def\Lcal{\mathcal{L}}  
\def\Rcal{\mathcal{R}}  
\def\Scal{\mathcal{S}}  
\def\conv#1#2#3#4#5{\(#1 \ast #2 \)_{#3}^{#4}\left[#5\right]}

\newif\ifshowWriterComment
\newcommand\writercomment[3]{\expandafter}

\writercomment{Shih-Hao}{SHT}{blue}
\writercomment{Lisa}{JSL}{green}

\def\fig#1{Fig.~\ref{fig:#1}}
\def\sec#1{Section~\ref{sec:#1}}
\def\tab#1{Table~\ref{tab:#1}}
\def\eqn#1{\eqref{eqn:#1}}

\def\mat#1{\begin{bmatrix}#1\end{bmatrix}}
\def\t{[t]}
\def\tn{[t+1]}
\def\st{{\rm s.t.}}
\def\OptConsSep{&&\quad}

\newcommand{\norm}[1]{\left\lVert#1\right\rVert}

\newcommand{\OptMin}[2]{
\begin{alignat}{2}
\min\ &\ #1 \nonumber \\
\st\ #2
\end{alignat}
}

\newcommand\OptCons[3]{
&\ #1
\ifx\\#2\\ \else \OptConsSep #2 \fi%
\ifx\\#3\\ \nonumber \else \label{eqn:#3} \fi%
}

\title{\LARGE \bf
SLSpy: Python-Based System-Level Controller Synthesis Framework
}

\author{Shih-Hao Tseng and Jing Shuang (Lisa) Li
\thanks{Shih-Hao Tseng and Jing Shuang (Lisa) Li are with the Division of Engineering and Applied Science, California Institute of Technology, Pasadena, CA 91125, USA.  Emails: {\tt\small \{shtseng,jsli\}@caltech.edu}}
}

\showWriterCommenttrue

\begin{document}

\maketitle
\thispagestyle{empty}
\pagestyle{empty}

\bstctlcite{IEEE_BSTcontrol}

\begin{abstract}

Synthesizing controllers for large, complex, and distributed systems is a challenging task. Numerous proposed methods exist in the literature, but it is difficult for practitioners to apply them -- most proposed synthesis methods lack ready-to-use software implementations, and existing proprietary components are too rigid to extend to general systems. To address this gap, we develop SLSpy, a framework for controller synthesis, comparison, and testing.

SLSpy implements a highly extensible software framework which provides two essential workflows: synthesis and simulation. The workflows are built from five conceptual components that can be customized to implement a wide variety of synthesis algorithms and disturbance tests. SLSpy comes pre-equipped with a workflow for System Level Synthesis (SLS), which enables users to easily and freely specify desired design objectives and constraints. We demonstrate the effectiveness of SLSpy using two examples that have been described in the literature but do not have ready-to-use implementations. We open-source SLSpy to facilitate future controller synthesis research and practical usage.

\end{abstract}
\section{Introduction}\label{sec:introduction}

Many of the systems we seek to control are large, complex, distributed, and multi-agent.
Controlling such systems is a nontrivial task, and there is a rich body of work surrounding this topic, with numerous proposed controller synthesis methods \cite{Nayyar2013, Bamieh2005, Rotkowitz2005, Ho1971, Bamieh2002, Mahajan2012}. Mostly, the methods deal with some mathematical program in the following form
\OptMin{
g(\xbf, \ubf)
}{
&\ \ubf = \Kbf(\xbf),\quad
(\xbf,\ubf) \in \Scal \nonumber
}
where $g$ is the objective, $\xbf$ the input signal, $\ubf$ the control signal, and $\Scal$ the feasibility constraints (including state dynamics, internal states, etc.). The synthesis methods then produce the desired controller map $\Kbf$.

Although mathematically rigorous, these synthesis methods are often inaccessible to engineers and control practitioners who want to test them out, customize them, or compare them with other methods.
For instance, many synthesis methods utilize frequency domain signals $\xbf$ to obtain controller $\Kbf$ and control signals $\ubf$, usually in the frequency domain as well. In practice, the raw input signal $x(t)$ the engineers have access to is in the time domain, and the control signal $u(t)$ should also be in time domain. Mapping the frequency domain controller $\Kbf(\xbf)$ to a time-domain implementation is not trivial. Furthermore, a recent study \cite{tseng2020deployment} shows that there exist multiple ways to implement the same frequency domain controller, each leading to different system properties. Thus, translating controller synthesis theories into implementable controllers can present significant challenges to control practitioners who are not theory experts.
From the users' perspective, it is more accessible if the synthesis methods work as a black box: given the system information and goals, derive a time-domain controller that takes the input $x(t)$ and returns the control $u(t)$. This can be realized via a software implementation.
Most methods described in the literature do not come with a ready-to-use software implementation. For those that do, the software may be proprietary and expensive and/or extend poorly to general systems.

Even when the software is available, it often comes in the form of toolboxes, which, though easy-to-use, are difficult to customize. Different toolboxes contain different sets of tools with different logical dependencies among them. Introducing new features involves fighting through these tangled dependencies.
Meanwhile, different toolboxes could contain similar components, e.g., linear system models, but implemented in different manners. Modifying such components could bog the user down in various logical threads behind the toolboxes and become an error-prone nightmare.

To avoid these hurdles and facilitate customization, we instead seek a \textit{framework} for controller synthesis. A framework is a structure that defines some conceptual components and dictates the dependencies among them -- the \emph{workflows}. To use it, the users provide customized component instances, which are then invoked by the framework to fulfill designated purposes of the workflows. Customization of components can be done via \textit{inheritance} in software, and customized components are easily used in conjunction with existing components thanks to the predefined workflows. Thus, frameworks provide a natural and easy way to extend and compare various algorithms and methods; they are used in domains such as robotics \cite{Lemaignan2015, Rohmer2013, Bruyninckx2013} and networking \cite{ns-3}.

In this paper, we introduce SLSpy: an open-source framework for discrete-time controller synthesis, which expedites the implementation and comparison of various synthesis methods. Although we provide SLSpy as a Python-based implementation, our proposed framework can work with components implemented in any other programming language (or even hardware) as long as the components obey the framework with an appropriate interface to exchange time-domain signals. The framework provides two key workflows -- synthesis and simulation -- and can accommodate any synthesis algorithm that follows the workflow described in \cite{tseng2020deployment}.

We demonstrate how SLSpy facilitates implementation and customization of synthesis methods through implementations of System Level Synthesis (SLS) and Input-Output Parametrization (IOP) \cite{wang2019system, anderson2019system, furieri2019input} in the framework. This is the first open-source implementation of IOP, and is the first Python-based implementation of SLS; we previously developed a MATLAB toolbox for SLS, SLS-MATLAB \cite{li2020sls}. SLSpy includes additional functionality for SLS not found in SLS-MATLAB, such as output feedback and LQG objectives, and as a framework, allows for much easier customization than SLS-MATLAB.

The main contributions of SLSpy are to introduce a framework for general discrete-time controller synthesis, and to create the first Python implementations of SLS and IOP with modularized objectives and constraints. We define and further motivate the necessity and extensibility of a framework, and describe the benefits of modularized objective and constraint design in \sec{motivation}. The architectures of the framework and the modules are described in \sec{architecture}. We then demonstrate the usefulness of SLSpy in \sec{example} through two examples: controller synthesis via IOP and output-feedback LQG. Finally, we summarize and list open questions and extensions for future work in \sec{conclusion}.
\subsection{Notation}
Let $\Rcal\Hcal_{\infty}$ denote the set of stable rational proper transfer matrices, and $z^{-1}\Rcal\Hcal_{\infty} \subset \Rcal\Hcal_{\infty}$ be the subset of strictly proper stable transfer matrices. Lower- and upper-case letters (such as $x$ and $A$) denote vectors and matrices respectively, while bold lower- and upper-case characters and symbols (such as $\ubf$ and ${\Phibf_\ubf}$) are reserved for signals and transfer matrices.
We use ${\Phi_u}[\tau]$ to denote the $\tau^{\rm th}$ spectral element of a transfer function ${\Phibf_\ubf}$, i.e., ${\Phibf_\ubf} = \sum\limits_{\tau=0}^{\infty} z^{-\tau} {\Phi_u}[\tau]$.
For simplicity, we write $\conv{x}{y}{lb}{ub}{t}$ as a shorthand notation for the discrete-time finite convolution $\sum\limits_{\tau = lb}^{ub} x[\tau]y[t - \tau]$.

\section{Motivation and Background}\label{sec:motivation}
We begin with the preliminaries of frameworks and System Level Synthesis (SLS). We describe the \emph{inversion of control} concept, which differentiates a framework from a toolbox/library, and explain why a framework is more extensible than a toolbox/library. Additionally, we illustrate how modularization can make SLS more accessible and customizable.

\subsection{Towards Framework: Inversion of Control}

Software is a crucial tool in the synthesis and design of controllers for large-scale systems. It allows synthesis algorithms to be distributed and used without the overhead of learning them in detail and implementing them. We summarize some state-of-the-art control and simulation software in \tab{state-of-the-art-methods}. Overall, most of the available tools for controller synthesis are proprietary, which are hard to extend and usually require license (and cost) to use. There are multiple open-source alternatives, but they do not directly serve the purpose of controller synthesis and are often domain-specific. We aim to develop an open-source general purpose software for controller synthesis.

\begin{table}[!t]
\centering
\caption{State-of-the-Art Control \& Simulation Software}
\label{tab:state-of-the-art-methods}
\renewcommand{\arraystretch}{1.25}
\begin{tabular}{|
@{}>{\centering}m{0.28\columnwidth}@{}||
@{}>{\centering}m{0.22\columnwidth}@{}|
@{}>{\centering}m{0.22\columnwidth}@{}|
@{}>{\centering}m{0.22\columnwidth}@{}|}
\hline
\multirow{2}{*}{Software} & \multicolumn{3}{c|}{Properties} \tabularnewline
\cline{2-4}
& Purpose & Open-Source & Framework \tabularnewline
\hline
\hline
Simulink\cite{Simulink} & general &  &  \tabularnewline
\hline
PSpice\cite{PSpice} & circuits &  &  \tabularnewline
\hline
SOSTOOLS\cite{prajna2002introducing} & optimization & $\surd$ & \tabularnewline
\hline
CVX\cite{CVX} & optimization & $\surd$ & \tabularnewline
\hline
ACADO\cite{Houska2011} & optimization & $\surd$ &  \tabularnewline
\hline
pyRobots\cite{Lemaignan2015} & robotics & $\surd$ &  \tabularnewline
\hline
SMACH\cite{Bohren2010} & robotics & $\surd$ &  \tabularnewline
\hline
V-REP\cite{Rohmer2013} & robotics & $\surd$ & $\surd$ \tabularnewline
\hline
TuLiP\cite{wongpiromsarn2011tulip,filippidis2016control} & temporal logic \\ planning & $\surd$ & \tabularnewline
\hline
SLS-MATLAB\cite{li2020sls} & controller \\ synthesis & $\surd$\footnote[1] & \tabularnewline
\hline
SLSpy & controller \\ synthesis & $\surd$ & $\surd$ \tabularnewline
\hline
\end{tabular}
\end{table}
\footnotetext[1]{Although SLS-MATLAB is open-source, it requires MATLAB, which is proprietary.}

\begin{figure}
\centering
\includegraphics[scale=1]{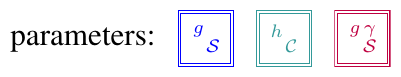}\\
\includegraphics[scale=1]{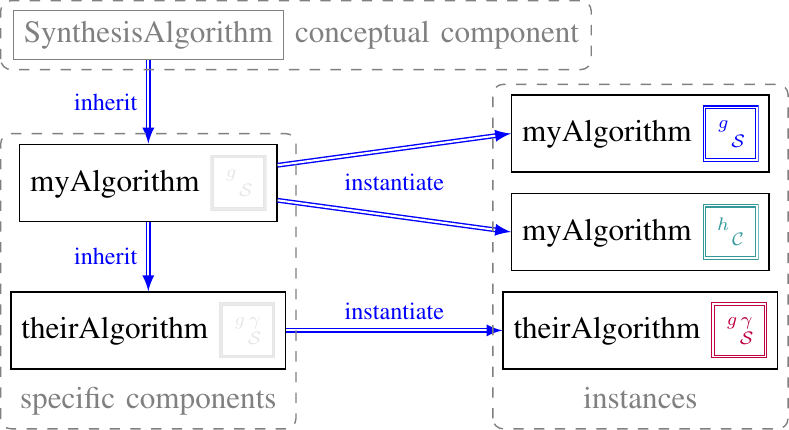}
\caption{An example of customization via inheritance. The first specific component (myAlgorithm) inherits from the conceptual component (SynthesisAlgorithm) and adds detailed behavior to create a usable algorithm. We can then use this specific component in code by instantiating it, i.e. by calling it with some specific parameters. Here, we have two instances of myAlgorithm, each with different parameters. Another specific component (theirAlgorithm) can inherit from myAlgorithm and add customized behavior of its own on top of the original algorithm.}
\label{fig:motivation-3tiers}
\end{figure}

There are several ways to pack algorithms/methods into software. Two major options are frameworks and libraries/toolboxes. A framework defines some workflows, which govern how various \textit{conceptual components} (e.g. system model, controller model) interact. The users can then customize the workflows of interest by defining \textit{specific components} (e.g. a linear system with system matrices A, B, C, D) from the conceptual components or other specific components via inheritance, a mechanism in software that allows a ``child'' component to include behavior from its ``parent'' component while also adding customized behavior of its own. Finally, the user \textit{instantiates} the component with some parameters (e.g. specific system matrices for some plant of interest) to obtain desired behavior. This is demonstrated in \fig{motivation-3tiers}. On the other hand, a library/toolbox consists of several functions/tools that perform specific actions. The users plan their workflows and choose the functions that fit in their schemes.

\vspace*{0.5\baselineskip}

\begin{figure}
\centering
\includegraphics[scale=1]{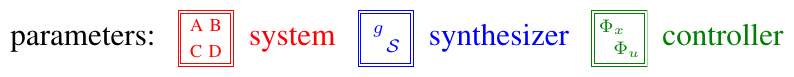}\\
\includegraphics[scale=1]{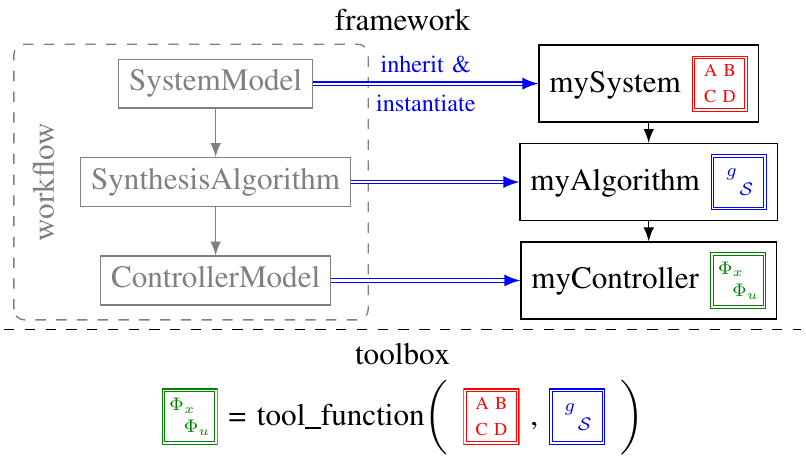}
\caption{Comparison between a framework and a toolbox. A framework pre-defines some workflow that users can follow. The users inherit and instantiate each conceptual component and execute the workflow. In contrast, a toolbox provides some tools that help the users obtain some results of interest. The users maintain the parameters and call the tools when needed. }
\label{fig:motivation-framework-vs-toolbox}
\end{figure}

A key property that distinguishes frameworks from libraries/toolboxes is the \emph{inversion of control} \cite{johnson1988designing, Inversion-of-Control} (also dubbed the Hollywood Principle -- ``Don't call us, we'll call you''). We illustrate this in the controller synthesis example shown in \fig{motivation-framework-vs-toolbox}.

To synthesize a controller, a possible workflow defined by a framework consists of three phases: establishing a system model, specifying the synthesis algorithm, and calculating the controller model. These are embodied by the conceptual components SystemModel, SynthesisAlgorithm, and ControllerModel. To use the framework, the user customizes the conceptual components to create the specific components myModel, myAlgorithm, and myController; then, the user instantiates the specific components with some parameters. The pre-defined workflow then feeds myModel into myAlgorithm to calculate the desired controller model (myController).

On the other hand, using a toolbox that contains some suitable tool (tool\_function), users maintain their system and synthesizer parameters and obtain the controller parameters via tool\_function. The two procedures differ in which entity controls the flow of the program; under a framework, the workflow is predetermined by the framework itself, whereas a toolbox allows the user to control the workflow. Therefore, control of the program is ``inverted'' from the user to the framework.

\subsection{Framework-Enabled Extensibility}

\begin{figure}
\centering
\includegraphics[scale=1]{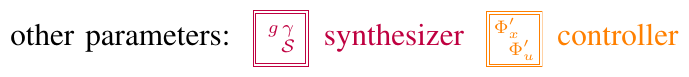}\\
\includegraphics[scale=1]{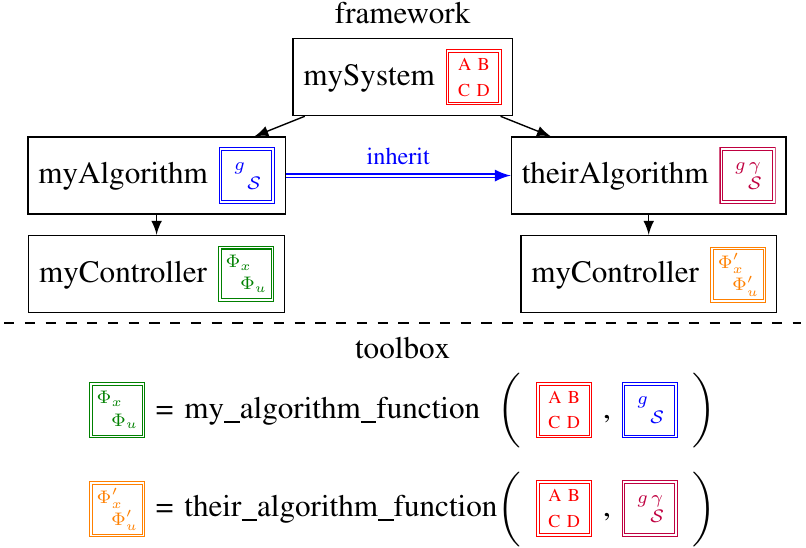}
\caption{Frameworks are highly extensible as we can reuse existing components in a new workflow instance. In this example, we are able to reuse the system model instance and controller model component (in different instances holding different parameters) while modifying the synthesis algorithm component via inheritance. In contrast, extending a toolbox requires directly modifying existing tool functions or introducing new tool functions, which is more involved.}
\label{fig:motivation-framework-extensibility}
\end{figure}

A framework's component-based architecture enables a user to easily customize select components as shown in \fig{motivation-3tiers}. Once this component has been customized, it is also easy to use alongside existing components, thanks to the framework's predefined workflow; little additional effort from the user is required to interface customized components with existing components. We demonstrate this in \fig{motivation-framework-extensibility}.

\fig{motivation-framework-extensibility} shows an example in which we want to customize some aspect of an existing synthesis algorithm. Given some existing specific components (myModel, myAlgorithm, myController), only the algorithm component (myAlgorithm) needs to be customized. We customize the existing synthesis algorithm to obtain the new synthesis component (theirAlgorithm). This new component outputs some new parameters that are easily fed into the existing controller model (myController) by the predefined workflow.

Consider the use case where we want to compare the new synthesis algorithm with the original one. As they both synthesize controllers for the same system, we only need one system instance. The two different controller algorithms then generate two sets of parameters, which are fed into two instances of the myController component for comparison. Overall, the framework allows us to reuse the myModel instance and the myController component.

As demonstrated in the above example, customization in a framework is less involved as a large portion of existing code can be reused, and the customized components can be easily embedded thanks to the framework's predefined workflow. Conversely, in a toolbox, one would need to directly edit the source code or introduce a new toolbox function, which are more involved and likely to introduce code repetition.

\subsection{Modularized System Level Synthesis}\label{sec:motivation-SLS}
Synthesizing an optimal controller for a networked cyber-physical system is challenging. The recently proposed System Level Synthesis (SLS) \cite{anderson2019system,wang2019system} method provides a solution for the following system:
\begin{align*}
x\tn =&\ A x\t + B_1 w\t + B_2 u\t, \\
\overline{z}\t =&\ C_1 x\t + D_{11} w\t + D_{12} u\t, \\
y\t =&\ C_2 x\t + D_{21} w\t + D_{22} u\t,
\end{align*}
where $x\t$ is the state, $w\t$ the noise, $u\t$ the control, $\overline{z}\t$ the regulated output, and $y\t$ the measurement at time $t$. SLS aims to synthesize a controller, the transfer function $\Kbf$ that maps the state $\xbf$ or the output $\ybf$ to the control $\ubf$, subject to some system-level objective $g$ and constraint $\Scal$. To do so, SLS introduces a new parametrization such that by solving
\OptMin{
g(\Phibf_\xbf,\Phibf_\ubf)
}{
\OptCons{
\mat{zI-A & -B_2}
\mat{{\Phibf_\xbf}\\
{\Phibf_\ubf}
}
=
I,
}{}{sf-constraint}\\
\OptCons{
{\Phibf_\xbf},{\Phibf_\ubf} \in z^{-1}\Rcal\Hcal_{\infty},
}{}{}\\
\OptCons{
\mat{{\Phibf_\xbf}\\
{\Phibf_\ubf}} \in \Scal,
}{}{}
}
for a state-feedback system and
\begin{subequations}
\OptMin{
g(\Phibf_{\xbf\xbf}, \Phibf_{\ubf\xbf}, \Phibf_{\xbf\ybf}, \Phibf_{\ubf\ybf})
}{
\OptCons{
\mat{zI-A & -B_2}
\mat{
{\Phibf_{\xbf\xbf}} & {\Phibf_{\xbf\ybf}} \\
{\Phibf_{\ubf\xbf}} & {\Phibf_{\ubf\ybf}} \\
}
=
\mat{I & 0},
}{}{of-constraint-1}\\
\OptCons{
\mat{
{\Phibf_{\xbf\xbf}} & {\Phibf_{\xbf\ybf}} \\
{\Phibf_{\ubf\xbf}} & {\Phibf_{\ubf\ybf}} \\
}
\mat{zI-A \\ -C_2}
=
\mat{I \\ 0},
}{}{of-constraint-2}\\
\OptCons{
{\Phibf_{\xbf\xbf}},{\Phibf_{\ubf\xbf}},{\Phibf_{\xbf\ybf}} \in z^{-1}\Rcal\Hcal_{\infty},{\Phibf_{\ubf\ybf}} \in \Rcal\Hcal_{\infty},
}{}{}\\
\OptCons{
\mat{
{\Phibf_{\xbf\xbf}} & {\Phibf_{\xbf\ybf}} \\
{\Phibf_{\ubf\xbf}} & {\Phibf_{\ubf\ybf}} \\
} \in \Scal,
}{}{}
}
\end{subequations}
for an output-feedback system, we can derive the controllers of the corresponding systems by
\begin{align*}
\text{state-feedback:} && \ubf =&\ \(\Phibf_\ubf \Phibf_\xbf^{-1}\) \xbf,\\
\text{output-feedback:} && \ubf =&\ \(\Phibf_{\ubf\ybf} - \Phibf_{\ubf\xbf}\Phibf_{\xbf\xbf}^{-1}\Phibf_{\xbf\ybf}\) \ybf.
\end{align*}
For simplicity, we denote by $\Phibf$ the set of SLS parameters, i.e., $\{ \Phibf_\xbf,\Phibf_\ubf \}$ or $\{ \Phibf_{\xbf\xbf}, \Phibf_{\ubf\xbf}, \Phibf_{\xbf\ybf}, \Phibf_{\ubf\ybf} \}$.

We refer the interested reader to \cite{wang2019system, anderson2019system} for details, where the motivation, derivation, and benefits of the SLS parametrization are thoroughly discussed.

A key feature of SLS is that it enforces the constraint $\Scal$ explicitly through the optimization. As such, it decouples the solving procedure from the structure of constraints. This entanglement greatly confined the capability of legacy methods, e.g., \cite{youla1976modern, lamperski2012dynamic, sabau2014youla}, to approach only certain constraints and systems. With SLS, we can now specify the constraints freely and let the corresponding convex program determine the feasibility. To facilitate the usage of SLS, we have developed and released the SLS-MATLAB toolbox \cite{li2020sls, GitHub-SLS-Caltech}. However, the toolbox does not easily extend to novel or custom SLS-based methods, which often require edits to the core code of the toolbox; here, again, we seek a \textit{framework} to enable customization.

\begin{figure}
\centering
\includegraphics[scale=1]{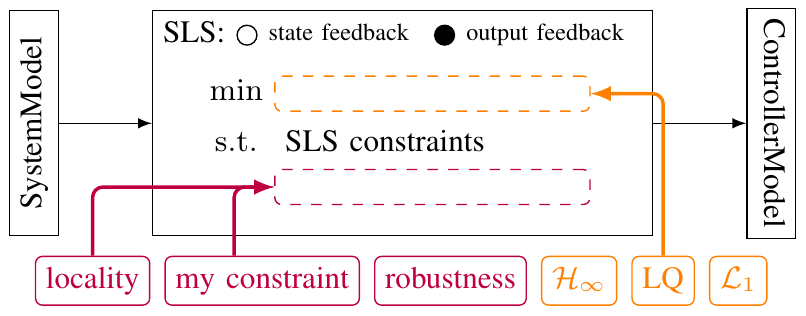}
\caption{We automate the synthesis process and modularize objectives and constraints so that users can focus on selecting, customizing, or even combining the modules according to their needs without learning and implementing the underlying theories of System Level Synthesis (SLS).}
\label{fig:motivation-SLS}
\end{figure}

From a practical perspective, users of SLS care more about obtaining a controller that meets their specifications than about the details of the underlying optimization. User-specified requirements on the controller correspond to objectives and constraints in the SLS problem. Motivated by this, we propose to automate the synthesis process and modularize the objectives and constraints as shown in \fig{motivation-SLS}. This allows the users to specify the synthesis type (state-feedback or output-feedback) and select, \textit{customize}, or even combine their desired objective and constraint modules. The framework then carries out the synthesis and generates the controller model for the users. Through modularization, we aim to make SLS more accessible to not only researchers but also control practitioners.

\section{Architecture}\label{sec:architecture}
We design SLSpy, a software framework for system-level controller synthesis. Our framework addresses the controller synthesis problem at the system level; component-wise details are omitted and the system is described by a map between its sensors and actuators.

The extensibility of a framework relies on the ability to customize components via inheritance.
For this reason, we implement our framework in Python, an objected-oriented language with good support for inheritance. An additional benefit of Python is that it is open-source and commonly used, which makes our framework more accessible. We remark that the concepts in this paper are not Python-specific; our framework can be implemented in any programming language that supports inheritance or some equivalent instantiation process.

Below, we illustrate the details of our framework and its SLS modules.

\subsection{Framework Overview}\label{sec:architecture-framework-overview}
To design a framework for system-level controller synthesis, we focus on two essential workflows: \emph{synthesis} and \emph{simulation}, as shown in \fig{workflows}. We further partition the workflows into five core conceptual components: \emph{SystemModel}, \emph{SynthesisAlgorithm}, \emph{ControllerModel}, \emph{NoiseModel}, and \emph{Simulator}. The synthesis workflow takes a SystemModel and synthesizes a desired ControllerModel. The simulation workflow allows users to verify the behavior of the resulting ControllerModel fed back to the SystemModel, and examine the impact of external disturbances from the NoiseModel.

We design the simulation workflow to lie in the time domain. As a result, all conceptual components should handle and produce time domain signals with the exception of SynthesisAlgorithm. This design decision allows the components to collaborate with real cyber-physical systems. For example, with appropriate hardware-software interfaces, a ControllerModel can generate control signals to control a real system; a physical controller can be tested with different SystemModels; a NoiseModel can serve as a noise generator for robustness tests of real systems. For flexibility of the synthesis workflow, we allow SynthesisAlgorithm to deal with the frequency domain. Overall, our framework maintains flexibility to accommodate as many future synthesis algorithms as possible.

We explain the functions of each component below:

\begin{figure}
\centering
\includegraphics{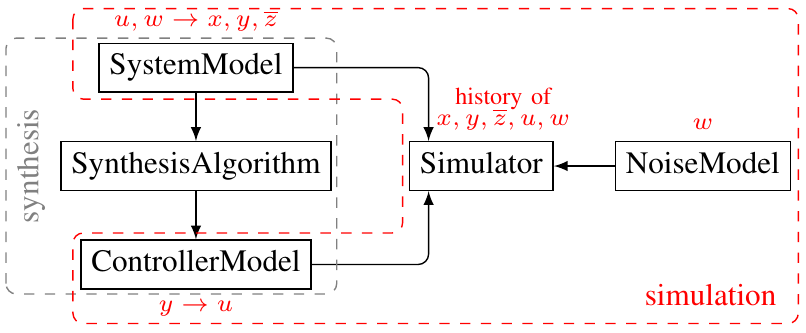}
\caption{The framework in SLSpy defines two workflows: synthesis and simulation. The two workflows consist of five core conceptual components: SystemModel, SynthesisAlgorithm, ControllerModel, NoiseModel, and Simulator. Besides the SysthesisAlgorithm, all components work on time domain rather than frequency domain, which allows them to be ported to real system directly.}
\label{fig:workflows}
\end{figure}

\renewcommand{\paragraph}[1]{\vspace{0.2cm}\textbf{#1}}

\paragraph{SystemModel}, interfered by noise $w$, takes control input $u$ to generate state $x$, measurement $y$, and regulated output $\overline{z}$. A SystemModel could have internal states, which allows it to model a wide range of systems, including general linear or nonlinear, time-invariant or time-varying ones.

\paragraph{ControllerModel} receives the measurement $y$ (which equals to $x$ under state-feedback schemes) to produce control input $u$.
ControllerModel is flexible to accommodate a wide range of parametrizations of the represented controller.
For example, we can parametrize the class of linear time-invariant controllers in ControllerModel by a direct map $\mathbf{K}$ from $y$ to $u$, the Youla parameter $\mathbf{Q}$ \cite{youla1976modern}, or the SLS parametrization \cite{anderson2019system,wang2019system}, which uses closed-loop maps from state disturbances and measurement error to the state and input (i.e., $\Phibf$). ControllerModel contains procedures that turn measurement $y$ into control $u$ in time domain according to the parameters.

\paragraph{SynthesisAlgorithm} takes a SystemModel and synthesizes a ControllerModel according to its design parameters and constraints.
In conventional toolboxes, such as TuLiP\cite{wongpiromsarn2011tulip,filippidis2016control} and SLS-MATLAB \cite{li2020sls}, the synthesis algorithm and controller model are often coupled. However, for the framework, we separate the two for better extensibility and reduced code duplication. For example, there are many ways to design a controller $\mathbf{K}$, including LQR and pole-placement methods. These are two separate synthesis algorithms corresponding to the same controller model; the code for the controller model would be duplicated if we combined the synthesis algorithm and controller model.

\paragraph{NoiseModel} models some disturbance or noise processes. A key design decision we made is to exclude NoiseModel from the synthesis workflow, and hence from the SystemModel and SynthesisAlgorithm.
Indeed, some synthesis algorithms may assume and target specific classes of noise (e.g. Gaussian noise), but we argue that the assumptions should be part of their synthesis parameters. We instead include NoiseModel in the simulation workflow, so that we can examine the system performance under different external disturbances. Of course, users are free to choose a NoiseModel that agrees with their assumptions.

\paragraph{Simulator} simulates time-domain system behavior for a specific system (SystemModel) and controller (ControllerModel) in the presence of noise (NoiseModel), and outputs the resulting history of state $x$, measurement $y$, regulated output $\overline{z}$, control $u$, and noise $w$. Users can then analyze the history, visualize it using our pre-written visualization tools, and compare simulations from different controllers.

We include the Simulator as a separate entity from the SystemModel for extensibility; when the system is known, the coupling between SystemModel and Simulator is apparent. However, for applications with plant uncertainty or related to system identification, the SystemModel used in design is not necessarily the same as the true system, which the Simulator uses. Separating the SystemModel and Simulator also allows us to test a single controller on a variety of systems.

\subsection{System Level Synthesis Modules}
\begin{figure}
\centering
\includegraphics{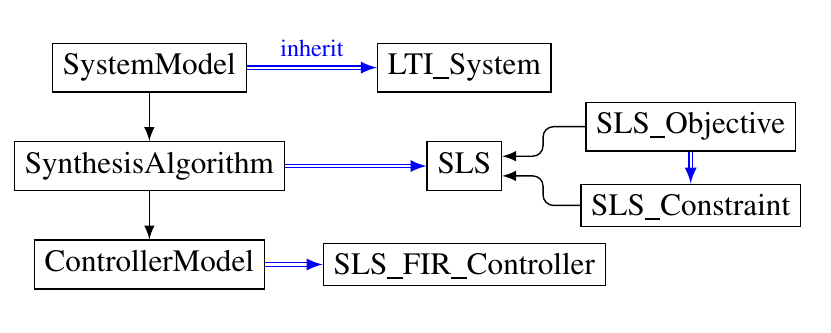}
\caption{SLSpy includes an implementation of SLS within the general framework, with SLS-specific objectives and constraints.}
\label{fig:sls-components}
\end{figure}

As illustrated in \sec{motivation-SLS}, SLS provides a new parametrization for both the formulation of the synthesis problem and the corresponding controller models. In addition to the conceptual components, SLSpy also includes some pre-defined specific components to facilitate usage of SLS, and to provide users with an example of how to use the framework. \fig{sls-components} shows how SLS is implemented within the SLSpy framework via inheritance; below, we describe the details of the implementation.

\begin{figure}
\centering
\includegraphics{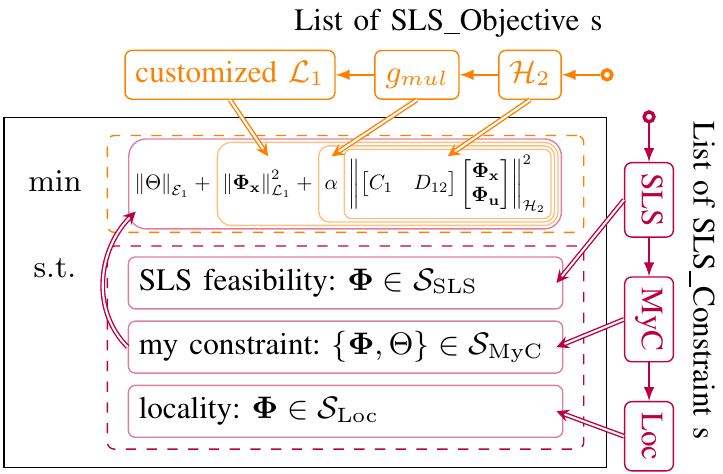}
\caption{SLSpy maintains two lists of user-selected SLS modules derived from the base classes SLS\_Objective and SLS\_Constraint. We then iterate through the lists to create the objective function and the list of constraints using the modules. Since SLS\_Constraint modules might introduce new variables and regularize them in the objective, SLS\_Constraint inherits from SLS\_Objective in our design.}
\label{fig:architecture-SLS-obj-cons}
\end{figure}

\paragraph{Constraints and Objectives}
Given an LTI\_System, the SLS algorithm formulates an optimization problem with some specified objective $g$ and constraint set $\Scal$. As stated in \sec{motivation-SLS}, we want to allow the user to specify arbitrary combinations of objectives and constraints. To this end, we include the \textit{SLS\_Objective} and \textit{SLS\_Constraint} base classes. We then maintain two lists, as shown in \fig{architecture-SLS-obj-cons}, to keep track of the user-selected modules, which are derived from SLS\_Objective and SLS\_Constraint. Below we explain how to combine those modules to form the corresponding SLS optimization problem.

A naive assumption for combining objectives is that the overall objective $g(\Phibf)$ is the sum of objective modules $g_i$, i.e., $g(\Phibf) = \sum\limits_{i} g_i(\Phibf)$. However, this is not the most general expression, and may lead to issues with more complex objectives. We instead make the more general assumption that the objective modules can modify the cumulative objectives from previous modules. Specifically,
\begin{align*}
g(\Phibf) = \dots g_3(\Phibf, g_2(\Phibf, g_1(\Phibf,0))).
\end{align*}
To demonstrate the flexibility of this structure, we consider the following objective as an example. Consider
\begin{align*}
g(\Phibf) = \alpha \norm{ \mat{C_1 & D_{12}} \mat{ \Phibf_{\xbf} \\ \Phibf_{\ubf} } }_{\Hcal_2}^2 + \norm{ \Phibf_{\xbf} }_{\Lcal_1}^2,
\end{align*}
which can be decomposed as
\begin{align*}
g(\Phibf) = g_3(\Phibf, g_2(\Phibf, g_1(\Phibf,0)))
\end{align*}
where
\begin{align*}
g_1(\Phibf,h) =&\ \norm{ \mat{C_1 & D_{12}} \mat{\Phibf_{\xbf} \\ \Phibf_{\ubf} } }_{\Hcal_2}^2 + h,\\
g_2(\Phibf,h) =&\ \alpha h,\\
g_3(\Phibf,h) =&\ \norm{ \Phibf_{\xbf} }_{\Lcal_1}^2 + h.
\end{align*}
Besides $\Phibf$ and $h$, each objective module can also take its own parameters to cover a larger class of objectives, e.g.,
\begin{align*}
g_{\Hcal_2}(\Phibf, C_1, D_{12}, h) = g_1(\Phibf,h), \ \ g_{mul}(\Phibf, \alpha, h) = g_2(\Phibf,h).
\end{align*}

We obtain $g$ by iterating through the SLS\_Objective list and performing function compositions. Correspondingly, SLS\_Objective must include a function for function composition.

Combining arbitrary constraints is trivial; we maintain a list of constraints and allow constraint modules to add to the list. Correspondingly, SLS\_Constraint should include a function that adds its constraint to the list. The core SLS feasibility constraints (\eqn{sf-constraint} for state feedback and \eqn{of-constraint-1}, \eqn{of-constraint-2} for output feedback) can be included as modules; they are generally applicable except in the case of robust SLS \cite{Matni2018}. Some SLS problems (e.g. robust SLS) are defined using a combination of new variables defined via equality constraints, and regularization terms on these new variables in the objective. In these cases, the constraint must be defined before the objective; for this reason, SLS\_Constraint inherits from SLS\_Objective.

\begin{figure}
\centering
\includegraphics{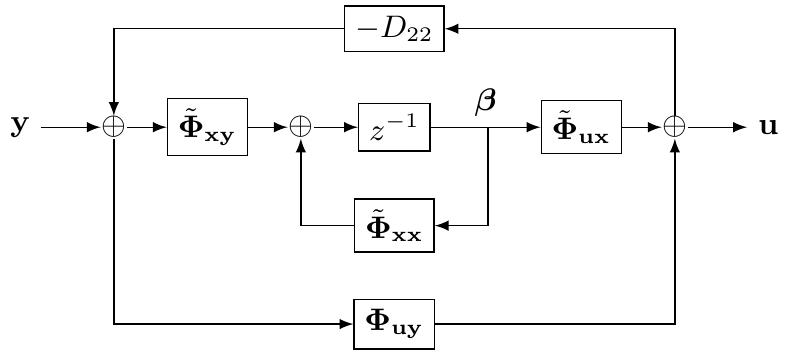}
\caption{Block-diagram realization of SLS output feedback controller, where $\tilde{\Phibf}_{\xbf\xbf} = z(I-z\Phibf_{\xbf\xbf})$, $\tilde{\Phibf}_{\ubf\xbf} = z \Phibf_{\ubf\xbf}$, and $\tilde{\Phibf}_{\xbf\ybf} = -z\Phibf_{\xbf\ybf}$.}
\label{fig:SLS-output-feedback-controller-diagram}
\end{figure}

\paragraph{Controllers}
SLS proposes controller realization in block diagrams, which are in the frequency domain.\footnote{We differentiate the ``realizations'' (block diagrams) from ``implementations'' (hardware/software architectures) of a controller according to \cite{tseng2020deployment}.}
The block-diagram realization of the SLS output feedback controller is shown in \fig{SLS-output-feedback-controller-diagram}. However, as described in \sec{architecture-framework-overview}, the ControllerModel requires functionality in the time domain. This necessitates the translation of \fig{SLS-output-feedback-controller-diagram} into time-domain equations for implementation.

\fig{SLS-output-feedback-controller-diagram} corresponds to the time-domain equations
\begin{align}
u[t] = (I + \Phi_{uy}[0] D_{22})^{-1}\(u'[t] + \Phi_{uy}[0] y[t]\)
\label{eqn:time-domain-of}
\end{align}
where the internal states are
\begin{align*}
u'[t] =&\ \conv{\Phi_{ux}}{\beta}{1}{T}{t-1} + \conv{\Phi_{uy}}{\overline{y}}{1}{T}{t-2}, \\
\beta[t+1] = &\ - \conv{\Phi_{xx}}{\beta}{2}{T}{t-2} - \conv{\Phi_{xy}}{\overline{y}}{1}{T}{t-1}, \\
\overline{y}[t] =&\ y[t] - D_{22} u[t].
\end{align*}

SLSpy implements the output-feedback SLS controller in time domain as defined in \eqn{time-domain-of}, as well as the state-feedback standard SLS controller in \cite{tseng2020deployment}.

\section{Examples}\label{sec:example}
Through the following examples, we demonstrate how SLSpy can help the user perform and study controller synthesis with ease. All codes used for the examples are available online at \cite{SLSpy}.

\subsection{Setup}
\begin{figure}
\centering
\includegraphics[scale=1]{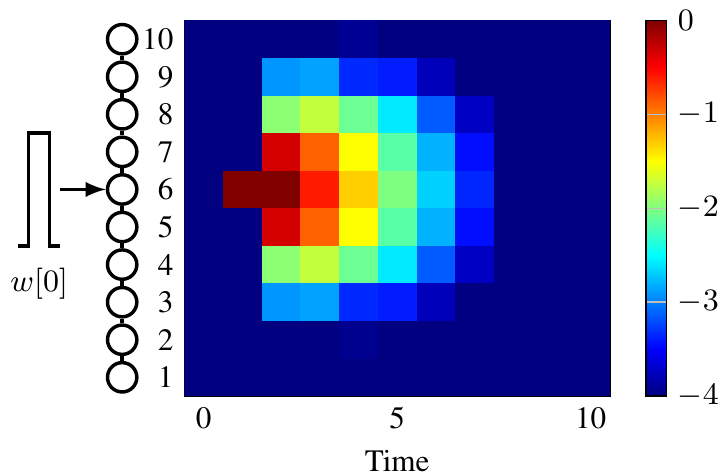}
\caption{In the examples, we consider a $10$-node fully-actuated chain-like system. At time $0$, an impulse disturbance $w[0] = 10$ hits its center, and we plot the time series of the quantities of interest in log scale. Each row of the plot shows the log-magnitude of the relevant signal (state, control, or measurement) of a single node over time; each column of the plot shows the log-magnitudes of all the nodes' signals at a single moment in time.}
\label{fig:example-setup}
\end{figure}

For all examples, we use a $10$-node fully-actuated chain-like system, as shown in \fig{example-setup}, with the following tridiagonal $A$ matrix:
\begin{equation}
A = \begin{bmatrix}
0.4 & 0.1 & 0 & \ldots & 0 \\
0.1 & 0.3 & \ddots & & \vdots \\
0 & \ddots & \ddots & \ddots & 0\\
\vdots & & \ddots & 0.3 & 0.1 \\
0 & \ldots & 0 & 0.1 & 0.4 \\
\end{bmatrix}
\end{equation}
The system is stable, with a spectral radius of $0.5$. We zero-initialize the system and disturb it at time $t=0$ with an impulse disturbance $w[0] = 10$. Under different controller models, we record the quantities of interest and plot their time series in log scale.

\subsection{Input-Output Parametrization}
The design of SLSpy framework allows the user to implement novel synthesis algorithms with ease. For example, a new parametrization -- Input-Output Parametrization (IOP) -- is proposed in \cite{furieri2019input} for the following system:
\begin{align*}
\ybf = \Gbf \ubf + \Pbf_{\ybf\wbf} \wbf,\quad
\zbf = \Pbf_{\zbf\ubf} \ubf + \Pbf_{\zbf\wbf} \wbf
\end{align*}
where $\ybf$, $\ubf$, and $\wbf$ are the measurement (system output), control, and noise, respectively. Given a transfer function $\Gbf$, IOP obtains the controller $\Kbf = \Ybf \Xbf^{-1}$ for $\ubf = \Kbf \ybf$ by solving
\OptMin{
\left\lVert \mat{
\Pbf_{\zbf\wbf} + \Pbf_{\zbf\ubf}\Ybf\Pbf_{\ybf\wbf}
} \right\rVert
}{
\OptCons{
\mat{
I & -\Gbf
}
\mat{
\Xbf & \Wbf\\
\Ybf & \Zbf
}
=
\mat{
I & 0
}
}{}{}\\
\OptCons{
\mat{
\Xbf & \Wbf\\
\Ybf & \Zbf
}
\mat{
-\Gbf\\
I
}
=
\mat{
0\\
I
}
}{}{}\\
\OptCons{
\Xbf, \Wbf, \Ybf, \Zbf \in \Rcal\Hcal_{\infty}.
}{}{}
}

\begin{figure}
\centering
\includegraphics{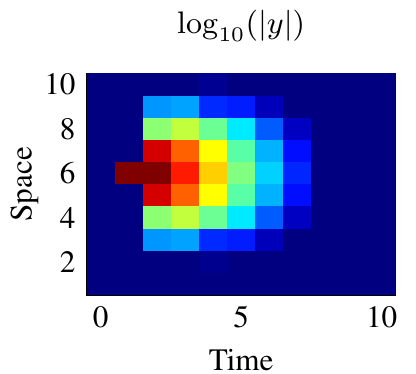}
\hfill
\includegraphics{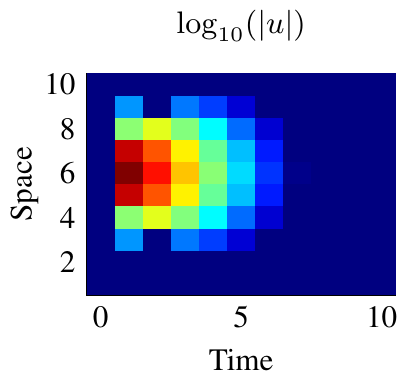}
\caption{We implement Input-Output Parametrization (IOP) using the SLSpy framework in only $282$ lines of code. The plot shows the system response to an impulse disturbance at the center of the chain; the IOP controller successfully stabilizes the system. Plots show the log magnitude of the measurement $y$ and the control $u$.}
\label{fig:IOP}
\end{figure}

We implement IOP in SLSpy with only $282$ lines of code, and we demonstrate the effectiveness of the IOP controller in \fig{IOP}. \fig{IOP} shows a simple example where the disturbance hits the center of a chain-like system of $10$ nodes. While the disturbance spreads, the IOP controller reacts and stabilizes the system.

\subsection{Output Feedback LQG}

\begin{figure}
\centering
\includegraphics{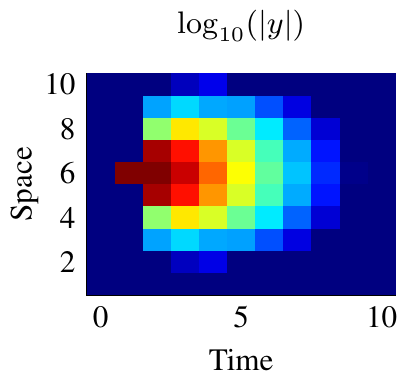}
\hfill
\includegraphics{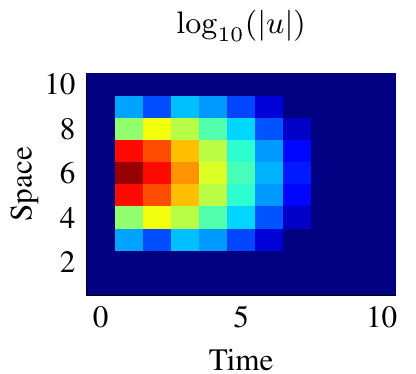}
\caption{We implement LQG using output feedback SLS, and simulate the system response to an impulse disturbance at the center of the chain, with no measurement noise. The system is successfully stabilized. Plots show the log magnitude of the measurement $y$ and the control $u$.}
\label{fig:LQG_SLS_noiseless}
\end{figure}

\begin{figure}
\centering
\includegraphics{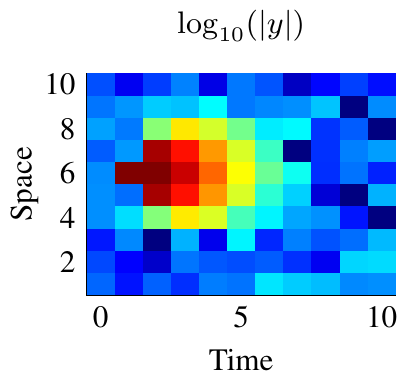}
\hfill
\includegraphics{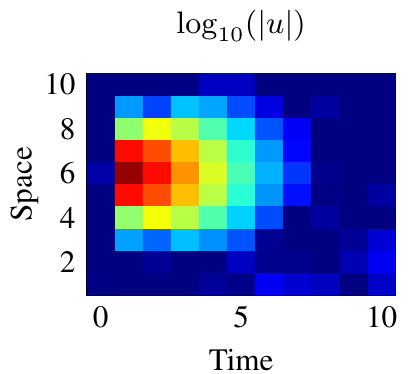}
\includegraphics{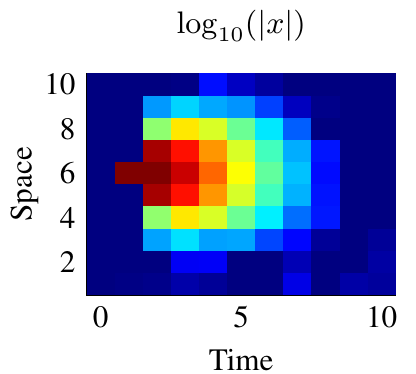}
\hfill
\caption{We implement LQG using output feedback SLS, and simulate the system response to an impulse disturbance at the center of the chain, with noisy measurements. Plots show the log magnitude of the measurement $y$, control $u$, and state $x$. The measurement noise is apparent in the plot of $y$; however, if we look at the plot of $x$, we see that the system is successfully stabilized.}
\label{fig:LQG_SLS_noisy}
\end{figure}

We implement an LQG controller with nonzero expected measurement noise. The SLS formulation of LQG can be found in \cite{Wang2016}. Since the framework decouples the expected noise (which is a parameter for controller synthesis) and the actual noise (which is used in the simulator), we can simulate the response of the controller to noises it was not designed for. We include a simulation of the LQG controller in a system with no measurement noise in \fig{LQG_SLS_noiseless}, and a simulation of the same controller in a system with measurement noise in \fig{LQG_SLS_noisy}.

Compared to the IOP controller, the LQG controller allows the disturbance to spread more in both time and space. Since the LQG controller expects measurement noise, it does not act as aggressively on sensor information as the IOP controller, which expects no measurement noise. A fairer comparison would be comparing the IOP controller with the LQR controller, and in that case, we find that the two controllers are identical, which matches the discussion in \cite{zheng2019equivalence}.

\section{Conclusion}\label{sec:conclusion}

We propose a software framework for controller synthesis and implement it as SLSpy in Python. Our framework serves as a platform for comparison of different discrete-time controller synthesis methods. We describe the architecture of the framework and its supported workflows, and use it to deploy modularized implementations of two synthesis methods that previously had no open-source implementations.

A direction for future work is exploring how additional optimization solvers and techniques can be incorporated into the framework. Currently, all objectives and constraints are directly specified in CVX syntax. One possible solution is the inclusion of a translator component between objectives/constraints and the solver.

\bibliographystyle{IEEEtran}
\bibliography{Test}

\end{document}